\documentclass[9pt,shortpaper,twoside,web]{ieeecolor2}
\usepackage{generic}
\usepackage{amsmath,amssymb,amsfonts}
\usepackage{algorithmic}
\usepackage{microtype}                 
\PassOptionsToPackage{warn}{textcomp}  
\usepackage{textcomp}                  
\usepackage{mathptmx}
\usepackage{amsmath}
\usepackage{times}                     
\usepackage{cite}                      
\usepackage{tabu}                      
\usepackage{booktabs}   
\usepackage{graphicx}
\usepackage{comment}
\usepackage{subcaption}
\usepackage{caption}
\usepackage{lipsum}
\def\BibTeX{{\rm B\kern-.05em{\sc i\kern-.025em b}\kern-.08em
    T\kern-.1667em\lower.7ex\hbox{E}\kern-.125emX}}
\begin{document}
\title{A Multi-Camera Optical Tag Neuronavigation and AR Augmentation Framework for Non-Invasive Brain Stimulation}
\author{Xuyi Hu, \IEEEmembership{Member, IEEE}, Ke Ma, Siwei Liu, Per Ola Kristensson, and Stefan Goetz, \IEEEmembership{Senior Member, IEEE}
\thanks{This paragraph of the first footnote will contain the date on 
which you submitted your paper for review. It will also contain support 
information, including sponsor and financial support acknowledgment. For 
example, ``This work was supported in part by the U.S. Department of 
Commerce under Grant BS123456.'' }
\thanks{Xuyi Hu is with the Department of Engineering, University of Cambridge, United Kingdom (e-mail: xh365@cam.ac.uk).}
\thanks{Ke Ma, Siwei Liu, Per Ola Kristensson, and Stefan Goetz are with the Department of Engineering, University of Cambridge, United Kingdom.}
}

\maketitle

\begin{abstract}
Accurate neuronavigation is essential for generating the intended effect with transcranial magnetic stimulation (TMS). Precise coil placement also directly influences stimulation efficacy. Traditional neuronavigation systems often rely on  costly and still hard to use and error-prone tracking systems. To solve these limitations, we present a computer-vision-based neuronavigation system for real-time tracking of patient and TMS instrumentation. The system can feed the necessary data for a digital twin to track TMS stimulation targets. We integrate a self-coordinating optical tracking system with multiple consumer-grade cameras and visible tags with a dynamic 3D brain model in Unity. This model updates in real time to represent the current stimulation coil position and the estimated stimulation point to  intuitively visualize neural targets for clinicians. We incorporate an augmented reality (AR) module to bridge the gap between the visualization of the digital twin and the real world and project the brain model in real-time onto the head of a patient. AR headsets or mobile AR devices allow clinicians to interactively view and adjust the  placement of the stimulation transducer intuitively instead of guidance through abstract numbers and 6D cross hairs on an external screen. The proposed technique provides improved spatial precision as well as accuracy. A case study with ten participants with a medical background also demonstrates that the system has high usability.
\end{abstract}


\section{Introduction}
\begin{figure}[!tb]
\centering
\includegraphics[width=0.6\linewidth]{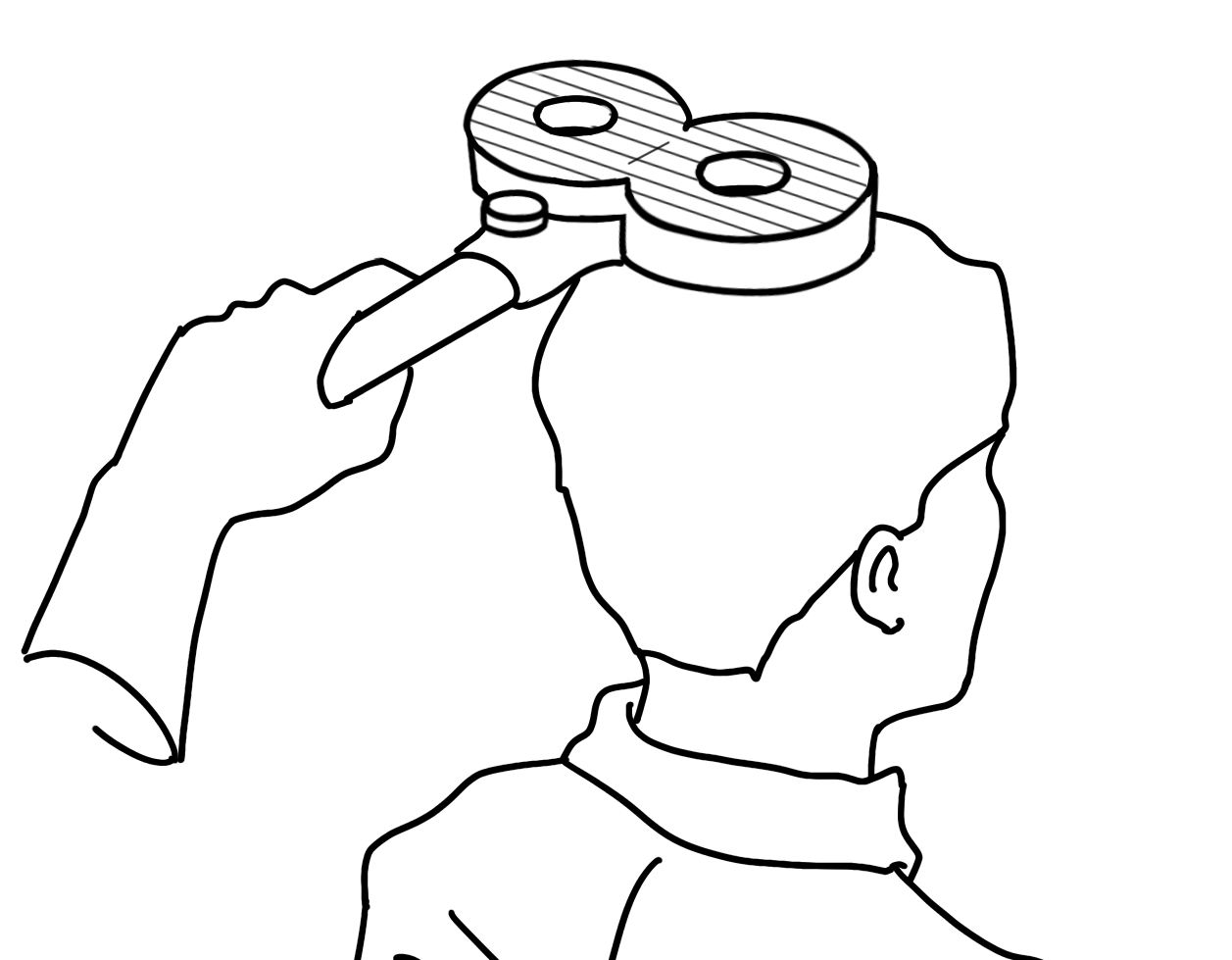}
\caption{\label{fig:TMS treatment}Illustration of TMS treatment. The TMS coil is placed on the patient’s head with millimeter accuracy above a specific neural target to target a certain circuit, e.g., often a control loop, for clinical therapy or brain research.}
\end{figure}

\maketitle
An exciting area of augmented reality (AR) is to better support therapy and research on the human brain. 
Transcranial magnetic stimulation (TMS) is a focal non-invasive technique to write signals into circuits of the brain through strong brief electromagnetic pulses across the intact scalp and skull \cite{barker1985non,wassermann2008oxford,wang2025single}. In contrast to deep brain stimulation \cite{perlmutter2006deep}, intracortical microstimulation \cite{flesher2016intracortical}, or epidural stimulation \cite{gerasimenko2008epidural,edgerton2011epidural}, TMS does not require any surgical implantation or even anesthesia. Instead, it works on awake patients.  A strong current pulse in a focal stimulation coil generates a local magnetic field~\cite{deng2017}. This magnetic field penetrates the scalp as well as skull with minimal distortion and induces a spatially confined electric field in the brain, which in turn depolarizes neurons to respond with so-called action potentials, that is, a neural signal. Figure~\ref{fig:TMS treatment} provides an illustration of a typical TMS treatment setup.

Certain pulse rhythms allow the modulation of a neural circuit, that is, they influence how this circuit or connection processes physiological signals~\cite{koehler2024closed,liu2025prior}. Accurate targeting of brain regions is essential for the success of TMS. Augmented reality would give great assistance in this often overwhelming process of coil positioning on a head without characteristic guidance structure by providing real-time visual guidance~\cite{leuze2018mixed}. By selectively stimulating targeted brain regions, TMS enables both therapeutic and research applications in neurology, psychiatry, neurosurgery, and cognitive neuroscience~\cite{terao2002basic,loo2005review,wang2023three}.

The TMS coil serves as a focal stimulation transducer. Its shape and location  determine the stimulation target~\cite{deng2017}. In a first approximation, the coil needs to be placed on top of a target~\cite{goetz2021oxford}. The simplest coil is a circular loop of wire that is placed tangentially to the scalp to induce an electric current in the opposite direction relative to the changing coil current. For example, if the coil current flows clockwise, the induced cortical current flows counterclockwise. The spatial distribution of this induced current mirrors the TMS coil’s shape but is somewhat blurred due to biological tissue properties~\cite{goetz2021oxford}. Different coil configurations, such as figure-eight coils \cite{sekino2015eccentric,gomez2018design}, enhance focal precision by concentrating the induced current at the intersection of the two loops. 

TMS is widely used in the treatment of neurological and psychiatric disorders, including depression~\cite{sonmez2019accelerated,perera2016clinical,wang2025mesoscale}, obsessive-compulsive disorder (OCD)~\cite{rodriguez1996transcranial}, and chronic pain~\cite{galhardoni2015repetitive,hamid2019noninvasive}. Cortical mapping and pre-surgical planning locates functions in the brain by sending test pulses and generates maps~\cite{lefaucheur2016value,krieg2017protocol}. 

In depression treatment, stimulation of the left dorsolateral prefrontal cortex (DLPFC) aims to up-regulate the mood-control network~\cite{pascual1996rapid,jahanshahi1998left}. For OCD, TMS targets hyperactive circuits in the cortico-striato-thalamo-cortical (CSTC) loop, reducing compulsive behaviors~\cite{li2016cortico}. In chronic pain management, stimulation of the motor cortex alters pain perception pathways, providing relief to patients with conditions such as neuropathic pain~\cite{lefaucheur2006use,leo2007repetitive} and fibromyalgia~\cite{marlow2013efficacy}. Beyond clinical applications, TMS is also used in neuroscience research to study brain function~\cite{hallett2000transcranial,ma2023correlating,liu2025iterative} and cognitive processes~\cite{luber2014enhancement,ma2022revised}.

Neuronavigation is essential in TMS treatment for achieving precise coil positioning, which directly affects stimulation accuracy and clinical efficacy. Misalignment of the TMS coil can lead to inconsistent results, reduced therapeutic effects, and unintended neural stimulation in adjacent circuits. Given the variability in individual brain anatomy, precise targeting is essential for optimizing treatment efficacy and reproducibility~\cite{herwig2001navigation,ma2025rethinking}. Without accurate guidance, TMS  fails to stimulate the intended cortical regions and the  neuromodulation effects are not reliable ~\cite{ruohonen2010navigated,ma2025optimal}.

Augmented reality (AR) has emerged as a promising interaction paradigm to bridge the gap between real-world environments and digital guidance. AR offers clinicians real-time spatial information that enhances precision and usability~\cite{frantz2018augmenting,schneider2021augmented}. By overlaying anatomical models or stimulation targets directly onto the patient’s head, AR-based systems allow clear and direct alignment of the TMS coil with cortical targets. This reduces the reliance on abstract numerical guidance or external displays. Such immersive feedback supports both novice and expert users in maintaining correct coil orientation and distance during stimulation. It thus promises to increase reproducibility and user confidence. Use cases include AR-assisted cortical mapping, pre-surgical planning, and neuronavigation in psychiatric treatment setups. They demonstrate the potential of AR to augment clinical workflows and improve treatment consistency~\cite{wang2023three,schneider2021augmented}.

\begin{figure*}[tb]
\centering
\includegraphics[width=0.95\textwidth]{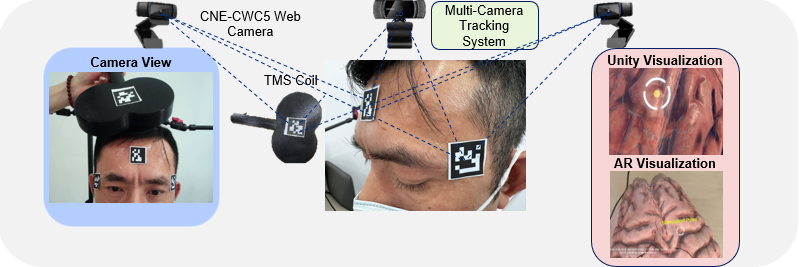}
\caption{The optical tag-based neuronavigation system combines multi-camera tracking with augmented reality (AR) visualization to improve transcranial magnetic stimulation (TMS) procedures taken from~\cite{hu2026opticaltagbasedneuronavigationaugmentation}. This system uses low-cost web cameras to track black-and-white optical fiducials on the TMS coil and patient for real-time tracking. Unity 3D visualization and AR overlays guide precise coil positioning and target identification.}
\label{fig:flowchart}
\end{figure*}

Various technologies have supported guidance of TMS coils in the past~\cite{goetz2021oxford}. Neuronavigation or so-called frameless stereotaxy often uses  optical tracking of retroreflective balls~\cite{gumprecht1999brainlab,ganslandt2002neuronavigation,muacevic2000accuracy,dockx2017accurate} or electromagnetic tracking of small transducers~\cite{hayhurst2009application} attached to fixed locations on the patient and the instruments. The stereo cameras of optical tracking are rather costly, typically operate in the infra-red spectrum, require pre-calibration, and triangulate the retroreflective balls only in a very limited volume from one perspective in which they require a free line of sight. Data is displayed on a screen, and operators correct their placement based on numbers, cross-hairs, and frequently some kind of 3D representation. The devices are complicated to use as the representation on a separate screen using numbers or a 3D image from a different perspective of the user is not easy to understand and therefore demands substantial training and practice. Further, the operating volume tends to be too small to be practical and, as a consequence, only a small part of the brain can be tracked from the camera's perspective at a time. Further, the optical or electromagnetic trackers are heavy and if not attached to bone screws, often move due to their weight. 

Computer vision (CV) promises a non-invasive and cost-effective solution for guiding TMS coil placement. In contrast to the heavy retro-reflective balls, trackers can be simple black-and-white patterns cost-effectively printed onto multiple stickers to be visible from many perspectives. Consumer-grade cameras have reached high resolutions and are readily available to allow multiple cameras from many view points to avoid shadow and perspective problems. Further, multiple cameras can compensate for abberation errors.

In this paper, we build on our novel approach presented in~\cite{hu2026opticaltagbasedneuronavigationaugmentation}. We present an AR-enabled neuronavigation system based on printed black-and-white tags, specifically AprilTags, integrated with AR for visualization to enhance accessibility and reduce cost of neuronavigation system (Figure~\ref{fig:flowchart}). This system leverages CV-based tracking to provide real-time, precise spatial localization of the TMS coil and the patient’s head. The primary contributions of this work are the following:
\begin{enumerate}
    \item We present a novel low-cost and easy to use tag-based multi-camera neuronavigation system for real-time tracking of TMS coil positioning within 5 mm accuracy at a total system cost of £60.  
    \item We introduce two additional evaluation methods to more comprehensively demonstrate that the proposed system is practical and competitive with existing, high-cost neuronavigation solutions.
    \item We modify the brain model and AR visualization to support both side-by-side display and spatially registered in-situ overlay directly on the patient’s head.
    \item We demonstrate the usability of the system in case study with ten participants with a medical background.
\end{enumerate}


\section{Conventional Neuronavigation Systems}
Optical neuronavigation systems use infrared stereo cameras to locate  reflective trackers mounted to the patient's head and tools  through triangulation to track the relative position and orientation of the TMS coil relative to the patient’s head. These systems can report the current position as well as the error relative to the intended target and guide the placement of the coil in real time~\cite{asfaw2024charting,marrone2024improving}. The primary advantage of optical tracking is its non-invasiveness and ability to dynamically adjust for head movement. 
However, optical neuronavigation systems have significant limitations ~\cite{goetz2021oxford, numssen2024quantification}. They require a constant unobstructed line of sight, and the cameras have a very limited operating range so that the camera cannot concurrently track all instruments, such as a coil on the right and the left hemisphere, or frontal and posterior, at the same time. Patients may also inadvertently and easily leave the covered range if they only moderately move or turn their head. Additionally, the rather heavy trackers, which are fixed to the head with tape, bands, or goggles, easily drift over time. As they define the coordinate system, a drift leads to systematic errors that are difficult to recognize, which demands regular recalibration. In addition, swaying trackers causes noise~\cite{matsuda2024robotic}. 

Electromagnetic (EM) neuronavigation provides an alternative to optical tracking systems and eliminates the need for a clear line of sight. Instead of cameras, EM systems generate a controlled electromagnetic field to track the head and instruments in real time~\cite{hayhurst2009application}. This technology is popular in neurosurgery, particularly in cases where optical tracking is obstructed or impractical. Studies have demonstrated  electromagnetic neuronavigation in pediatric neuro-endoscopic surgery, where its ability to function without direct visibility improves instrument maneuverability and procedural accuracy~\cite{choi2013usefulness}. However, electromagnetic systems are sensitive to metallic objects in the room or in devices, such as TMS coils, which can interfere with the magnetic field and introduce localization errors~\cite{behari2002neuronavigation}.


Commercial neuronavigation systems typically achieve sub millimeter accuracy~\cite{de2024precision, yu2024clinical}, but they require significant financial 
investment—often between \$30{,}000 and \$100{,}000 depending on vendor and configuration~\cite{sulangi2024neuronavigation, asfaw2024charting}. This corresponds to an accuracy-cost ratio exceeding £20000 per mm$^{-1}$. These systems are also complex to operate, demanding extensive training and integration into clinical workflows~\cite{wiedemann2021markerless}. In contrast, advances in consumer-grade cameras now offer high-resolution, high-frame-rate video capture, presenting a cost-effective alternative for optical tracking~\cite{wang2019pose, choi2018robust}.

While prior studies have explored coil tracking for TMS, our proposed system introduces several notable advancements. Washabaugh et al.~\cite{washabaugh2016low} developed a retroreflective-marker coil tracking system for TMS, which, although achieving acceptable accuracy and repeatability, suffers from several practical drawbacks. The absence of intuitive operator feedback limits usability, while the retroreflective markers themselves are not trivial in cost, and clusters of reflective spheres are needed for system calibration. Lin et al.~\cite{lin2019trajectory} presented a robotic trajectory control system for coil alignment; however, their approach depends on expensive hardware, involves a complex setup, and similarly lacks real-time, user-centered visual guidance. The reliance on a fixed robotic arm and laboratory-grade setup restricts portability. Such systems are unsuitable for resource-limited clinics or mobile applications. In addition, any drift in force sensing or delay in adaptive control could lead to patient discomfort or unstable positioning. Leuze et al.~\cite{leuze2018mixed} focused on mixed-reality visualization to support intuitive TMS coil guidance, but their approach remains limited to brain visualization and does not address stimulation target localization. In contrast, our method eliminates the need for robotic infrastructure by offering a lightweight, operator-guided solution that integrates real-time visual augmentation to facilitate intuitive and accurate coil positioning.

\section{Tag-Based Multi-Camera Tracking System}
\subsection{Tracking Algorithm}
\begin{figure}[!tb]
\centering
\includegraphics[width=0.8\linewidth]{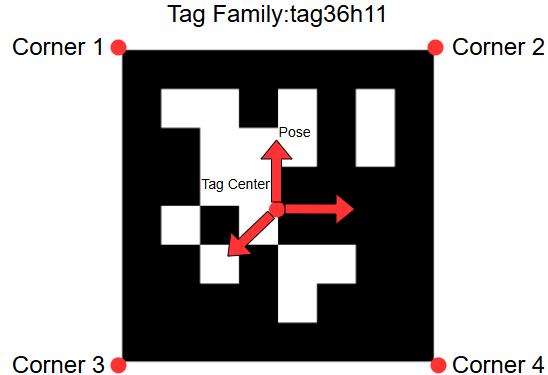}
\caption{\label{fig:AprilTag}Representation of the planar black-and-white tag family (AprilTag tag36h11). The tag's four corner points are labeled for accurate pose estimation. A coordinate frame at the tag center indicates its orientation and position in space, which can be derived from the corner positions even from a single camera perspective.}
\end{figure}

Our system uses regular rectangular black-and-white fiducial patterns from the AprilTag family~\cite{olson2011apriltag,wang2016apriltag}. It consists of high-contrast square patterns that encode unique identification numbers. Further, the sharp edges and the fundamental rectangular grid enable accurate estimation of location and orientation. The tracking algorithm extracts these markers from camera images, decodes their identities, and determines their 3D position and orientation relative to the camera through perspective transformation. Figure~\ref{fig:AprilTag} illustrates the AprilTag square pattern information.

In contrast to traditional reflective-marker-based frameless stereotaxy systems, the optical tags offer high robustness to lighting variations, occlusions, and motion blur. The algorithm first detects candidate quadrilateral regions through edge filtering and subsequently performs homography-based pose estimation to extract the marker’s position in 3D space. Once identified, multiple camera views can be integrated to improve spatial accuracy and reduce tracking errors.

Tracking based on optical tags with patterns in the visible spectrum provides several advantages over alternative tracking methods. It is computationally efficient and can operate without purpose-designed hardware and instead exploit the latest high-quality low-cost consumer-grade multi-camera and even monocular-camera systems. 

\subsection{Multi-Camera Setup}
To enable real-time 360 degree tracking of the TMS coil and patient’s head, we use a multi-camera system. This setup consists of three strategically positioned cameras for continuous spatial localization and minimizing occlusions. Each camera was individually calibrated offline to estimate intrinsic parameters using standard camera calibration procedures.
A shared world coordinate frame was established by combining the fixed geometry of the calibrated cameras with the known spatial arrangement of the head-mounted AprilTag markers.
This implicit world anchor enables consistent multi-camera pose estimation without requiring external tracking infrastructure.

The multi-camera system operates by capturing synchronized frames, where each camera independently detects AprilTag markers in its respective view.  We use computer vision to integrate positional data from all three cameras to calculate an accurate six-degrees-of-freedom (6 DoF) transformation for both the TMS coil and the head. The multi-camera setup increases accuracy and ensures that even if one camera loses track of a marker due to motion or occlusion, the system can still determine the correct coil position using data from the remaining cameras.

This approach enhances tracking robustness and ensures continuous real-time monitoring of TMS coil movement. Unlike traditional optical neuronavigation systems, which often require expensive hardware and calibration-intensive setups, this low-cost AprilTag based multi-camera system offers a scalable and adaptable solution for both clinical and research applications. The setup also supports dynamic head movement compensation, making it particularly useful for real-world applications where patient motion needs to be accounted for during TMS treatment sessions.


\subsection{Optical Target Estimation}
Accurate estimation of the target position is essential for precise TMS coil placement in neuronavigation. We use a multi-camera tracking system with AprilTag fiducial markers to estimate the transformation of the patient’s head and TMS coil. Five markers serve as sources for coordinate estimation, with four assigned to head tracking and one dedicated to coil tracking. The tag detection algorithm identifies the 2D pixel coordinates of the tag corners in each camera view, which are then processed to estimate the 3D position of the tags relative to each camera using intrinsic camera parameters and distortion coefficients. We construct a 3D world coordinate frame by combining the relative positions of the flat markers on the patient’s head with the fixed positions of the three calibrated cameras. This framework allows us to account for continuous motion introduced by patient head movement and operator hand tremor during the TMS procedure for accurate real-time tracking under naturalistic conditions. Therefore, even if some tags are occluded during operation, the system maintains stable target estimation due to redundant multi-camera observations and marker distribution.

\subsubsection{Image Formation}
In computer vision, image formation describes the process of mapping 3D world coordinates onto a 2D image plane using cameras. Given a 3D point \( P(U, V, W) \) in world coordinates, its corresponding 3D position in the camera coordinate system can be computed using a rigid body transformation consisting of a rotation matrix $\mathbf{R}$ and a translation vector $\mathbf{t}$:

\begin{equation}
\begin{bmatrix} X \\ Y \\ Z \end{bmatrix} = \mathbf{R} \begin{bmatrix}U \\ V \\ W \end{bmatrix} + \mathbf{t},
\end{equation}

The transformation can be rewritten in homogeneous coordinates as a four-vector equation

\begin{equation}
\begin{bmatrix} X \\ Y \\ Z \end{bmatrix} =
\begin{bmatrix}r_{00} & r_{01} & r_{02} & t_{x}\\
r_{10} & r_{11} & r_{12} & t_{y}\\
r_{20} & r_{21} & r_{22} & t_{z}\\
\end{bmatrix}
\begin{bmatrix}U \\ V \\ W \\ 1 \end{bmatrix}\!,
\end{equation}
where $r_{ij}$ defines the orientation of the world coordinate system relative to the camera coordinate system and $t_i$ represents the position of the world origin relative to the camera coordinate system.

\begin{figure}[t]
\centering
\includegraphics[width=1\linewidth]{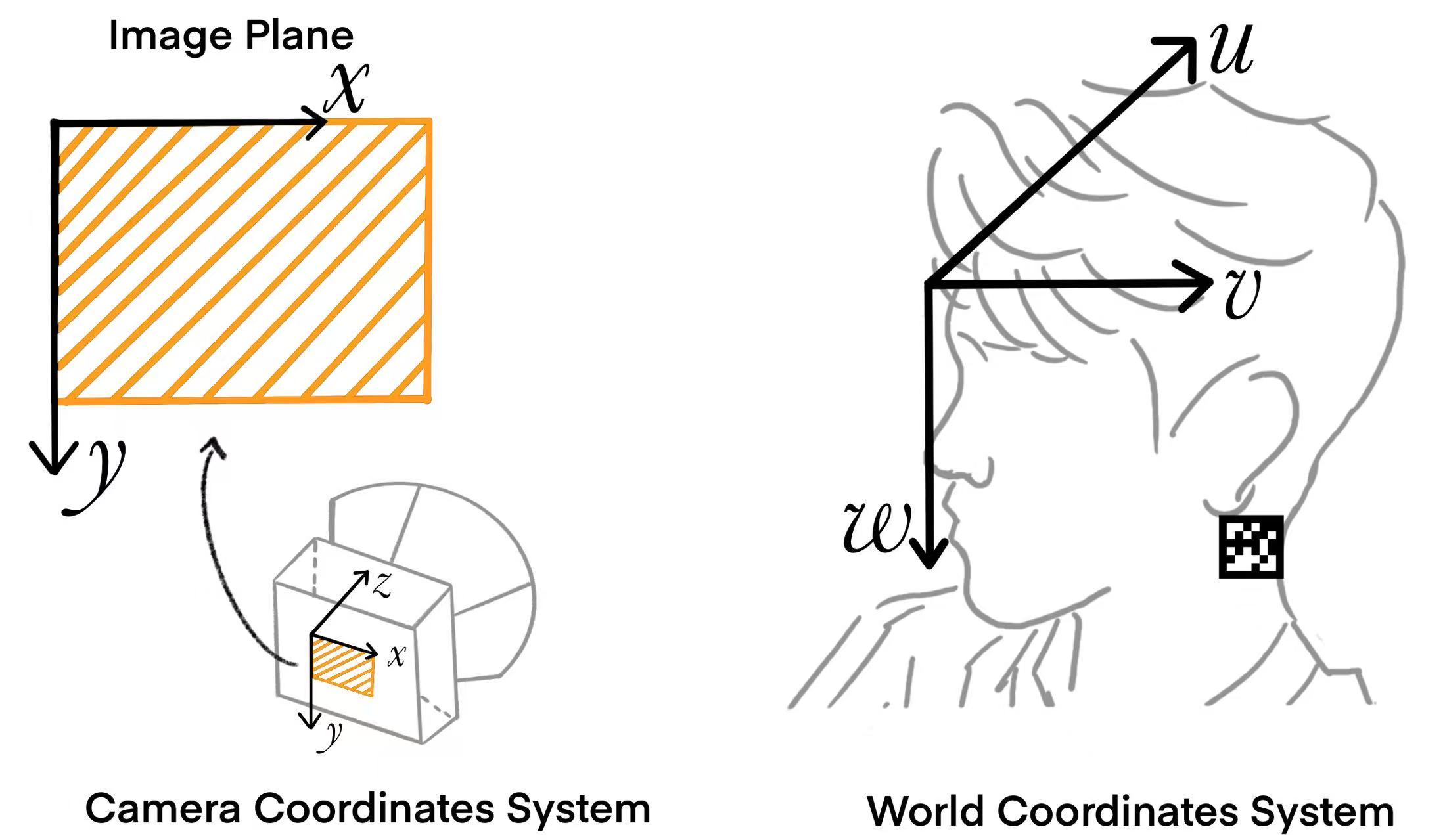}
\caption{\label{fig:coordinate system}Relationship between the image plane (2D pixel space), the camera coordinate system (3D space), and the world coordinate system (anchored to the patient via an fiducial marker). These transformations allow accurate localization of the TMS coil in the patient's head frame for real-time AR visualization and navigation.}
\end{figure}

\begin{figure*}[tb]
  \centering
  \begin{subfigure}[b]{0.45\textwidth}
    \centering
    \includegraphics[height=0.6\textwidth]{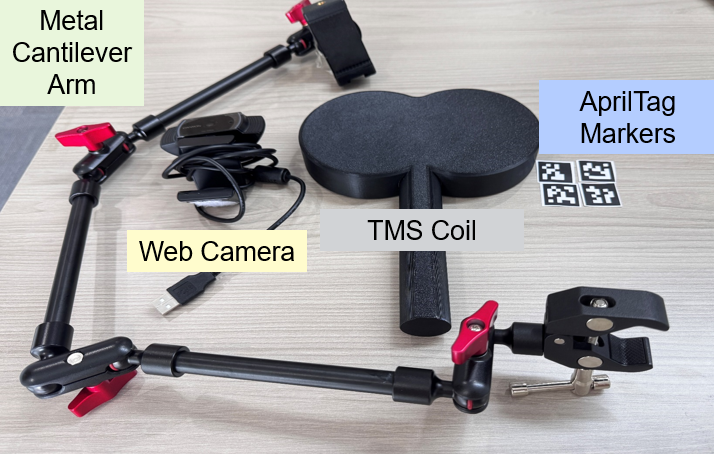}
    \caption{Tools and targets}
    \label{fig:pos1}
  \end{subfigure}
  \begin{subfigure}[b]{0.45\textwidth}
    \centering
    \includegraphics[height=0.6\textwidth, width=0.9\textwidth]{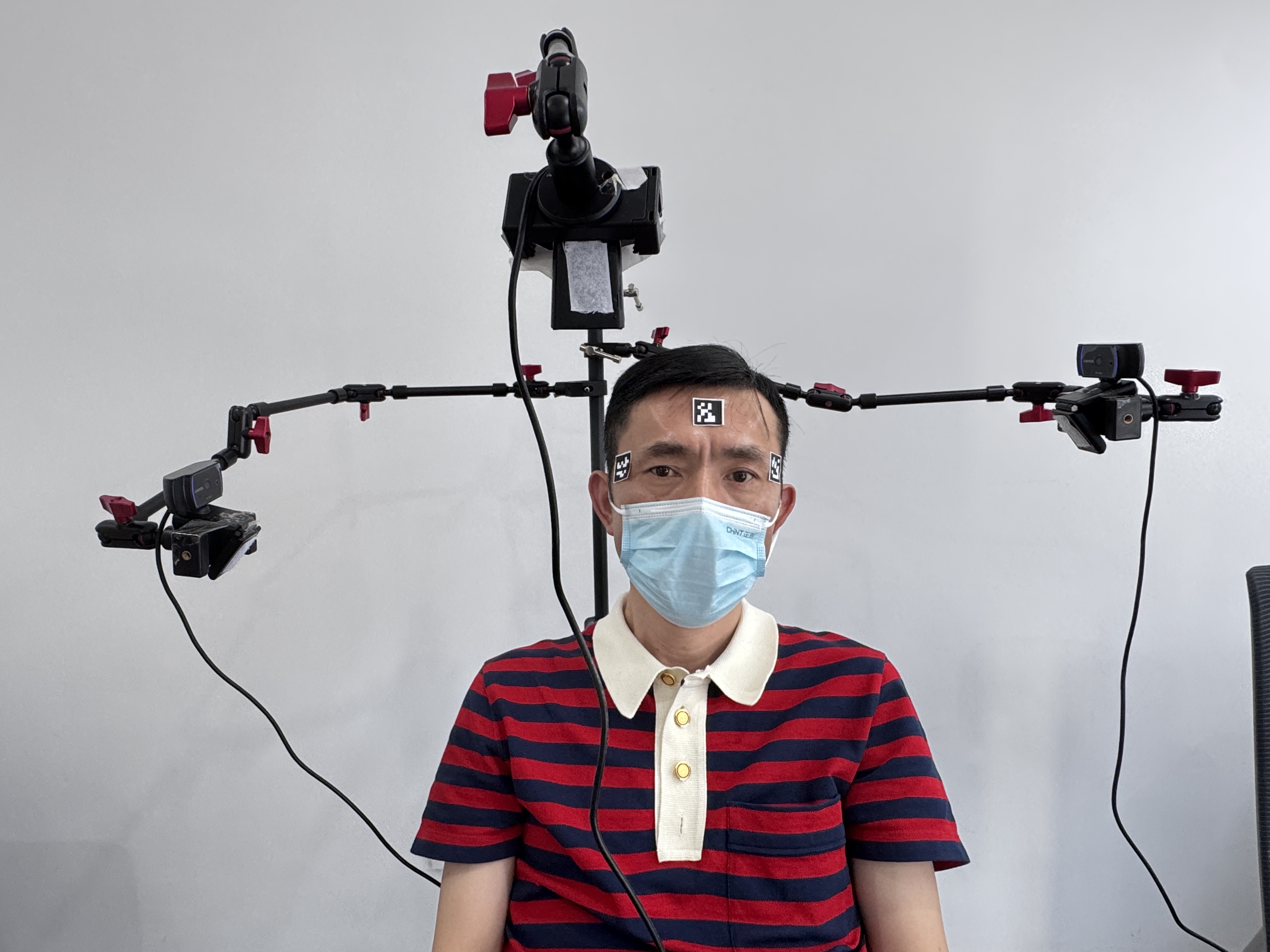}
    \caption{Multi-Camera TMS navigation workstation}
    \label{fig:pos5}
  \end{subfigure}
  \caption{(a) Components and (b) multi-camera TMS neuronavigation system during experiment taken from~\cite{hu2026opticaltagbasedneuronavigationaugmentation}. The system consists of a TMS coil, a multi-camera tracking setup for real-time spatial localization, and flat optical markers attached to the tracked objects and the head for precise position and orientation detection. The cameras provide synchronized tracking data and translate all components into a unified coordinate frame.}
  \label{fig:pipeline}
\end{figure*}

\subsubsection{Direct Linear Transform}
In pose estimation, we aim to determine the rotation and translation that align a known set of 3D points to their corresponding 2D projections. The transformation from 3D camera coordinates to 2D image coordinates follows the camera projection equation
\begin{equation}
\begin{bmatrix} x \\ y \\ 1 \end{bmatrix} = s
\begin{bmatrix} f_x & 0 & c_x \\ 0 & f_y & c_y \\ 0 & 0 & 1 \end{bmatrix}
\begin{bmatrix} X \\ Y \\ Z \end{bmatrix}
=s\mathbf{K}\begin{bmatrix} X \\ Y \\ Z \end{bmatrix},
\label{eq:world_to_camera}
\end{equation}
where \( f_x, f_y \) are the focal lengths in the \( x \) and \( y \) directions, \( c_x, c_y \) define the optical centre of the camera, \( s \) is depth scale factor and \(\mathbf{K}\) is the intrinsic camera matrix.

\subsubsection{Pose Estimation}
Once a fiducial marker is detected, the  algorithm provides the 2D image coordinates of the tag corners. Then, the camera projection matrix as defined in Equation~\ref{eq:world_to_camera}  establishes 3D-to-2D correspondences. Given a set of known 3D points in the tag’s coordinate frame and their corresponding 2D projections in the image plane, we can write
\begin{equation}
\begin{bmatrix}
x_i \\[4pt]
y_i \\[4pt]
1
\end{bmatrix}
=
\mathbf{K} \Bigl(
   \mathbf{R}
   \begin{bmatrix}
   X_i \\[3pt]
   Y_i \\[3pt]
   Z_i
   \end{bmatrix}
   \;+\;
   \mathbf{t}
\Bigr),
\end{equation}
where $\mathbf{R}$ is the rotation matrix, and $\mathbf{t}$ is the translation vector.

Given a set of known 3D landmarks (e.g., AprilTag marker corners) and their corresponding 2D image projections, we estimate the camera pose $\mathbf{R}$ and $\mathbf{t}$ by solving 
\begin{equation}
\min_{\mathbf{R}, \mathbf{t}} \sum_{i} \left\|\begin{bmatrix} x_i \\ y_i \\ 1 \end{bmatrix} - \pi \left(\mathbf{K} \biggl( \mathbf{R} \begin{bmatrix} X_i \\ Y_i \\ Z_i \end{bmatrix} + \mathbf{t}  \biggr)\right)\right\|^2\!\!{},
\label{eq:pose estimation}
\end{equation}
where $\pi(\cdot)$ is the camera projection function.

\subsection{Gaussian Error Combination for Increased Precision}



After Equation \ref{eq:pose estimation}, the reprojection error of camera pose $(\mathbf{r}, \mathbf{t})$ is estimated as follows:
\begin{equation}
e_{\text{proj}}
=
\frac{1}{N} 
\sum_{i=1}^{N}
\Bigl\|\,
\mathbf{x}_i
\;-\;
\widehat{\mathbf{x}}_i
\Bigr\|,
\label{eq:2D reprojection error}
\end{equation}
where $\mathbf{x}_i \in \mathbf{R}^2$ is the $i$-th corner’s detected pixel coordinate, and $\widehat{\mathbf{x}}_i$ is that corner’s projected pixel coordinate given the estimated pose.  This quantifies how well the solution $(\mathbf{R}, \mathbf{t})$ matches the observed 2D corners.

The uncertainty $\sigma_{\text{distance}}$ is derived from the re-projection error follows
\begin{equation}
\begin{aligned}
 \sigma_{t_x} &= e_{\text{proj}}\,\times \frac{t_z}{f_x},\\
 \sigma_{t_y} &= e_{\text{proj}}\,\times \frac{t_z}{f_x},\\
 \sigma_{t_z} &= e_{\text{proj}}\,\times \frac{t_z}{\sqrt{f_x^2 + f_x^2}},
\end{aligned}
\end{equation}
where $f_x$ is the camera’s focal length in pixels and $t_z$ is the depth in the camera coordinate system. 

The standard deviation for each camera’s distance estimate follows
\begin{equation}
\sigma_{\text{distance}}
=
\sqrt{
\Bigl(\frac{t_x}{d}\sigma_{t_x}\Bigr)^2
 +
\Bigl(\frac{t_y}{d}\sigma_{t_y}\Bigr)^2
 +
\Bigl(\frac{t_z}{d}\sigma_{t_z}\Bigr)^2
}.
\quad
\end{equation}

When multiple cameras detect the same tag, each camera $j$ provides distance $\mathbf{d}$ and standard deviation $\mathbf{\sigma_{d_j}}$. We combine the distance estimates statistically and consider their variability:
\begin{equation}
d_{\mathrm{fused}}
=
\frac
{
\sum_{j=1}^{m} \frac{d_j}{\sigma_{d_j}^2}
}
{
\sum_{j=1}^{m} \frac{1}{\sigma_{d_j}^2}
}
,\quad
\sigma_{\mathrm{fused}}
=
\sqrt{
\frac{1}{\sum_{j=1}^m \frac{1}{\sigma_{d_j}^2}}
}.
\end{equation}

Consequently, measurements from more reliable cameras (small $\sigma_{d_j}$) dominate the weighted average, and the resulting fused standard deviation $\sigma_{\mathrm{fused}}$ reflects the combined confidence. This algorithm is a Gaussian weighting scheme that yields an overall final distance estimate and uncertainty suitable for robust multi-camera tracking of the TMS coil or other tags.

\section{Experimental Setup}
\subsection{Setup}
Each camera is mounted at a distinct angle to ensure 3D coverage of the workspace. One camera is placed frontally to observe the head and coil face-on, while the other two are positioned laterally so that all optical tags on both the coil and the patient's head remain visible. Four different tag36h11 markers (AprilTag family), each measuring 24 × 24 mm, were attached to the head, three positioned on the front face and one on the back. Of the front-facing markers, two are placed near the cheekbones, where facial muscle movement is minimal, ensuring that the markers remain relatively stable during TMS treatment. The additional markers on the forehead and back of the head provided complementary reference points. They enable robust and reliable head pose estimation throughout the procedure. An additional fiducial marker is affixed to the center of TMS coil, enabling precise tracking of its spatial position and current simulation point. This multi‐view configuration allows real‐time pose estimation by fusing data across cameras, mitigating issues like marker occlusion or limited field of view from a single perspective.


We use three commercial cameras (CANYON CNE-CWC5), each with a $1920 \times 1280$ resolution and a 65° wide-angle horizontal field of view, for experiments. Each camera is priced at £21. The implementation details of the experiment are presented in Figure \ref{fig:pipeline}.

\begin{figure}[t]
\centering
\includegraphics[width=0.8\linewidth]{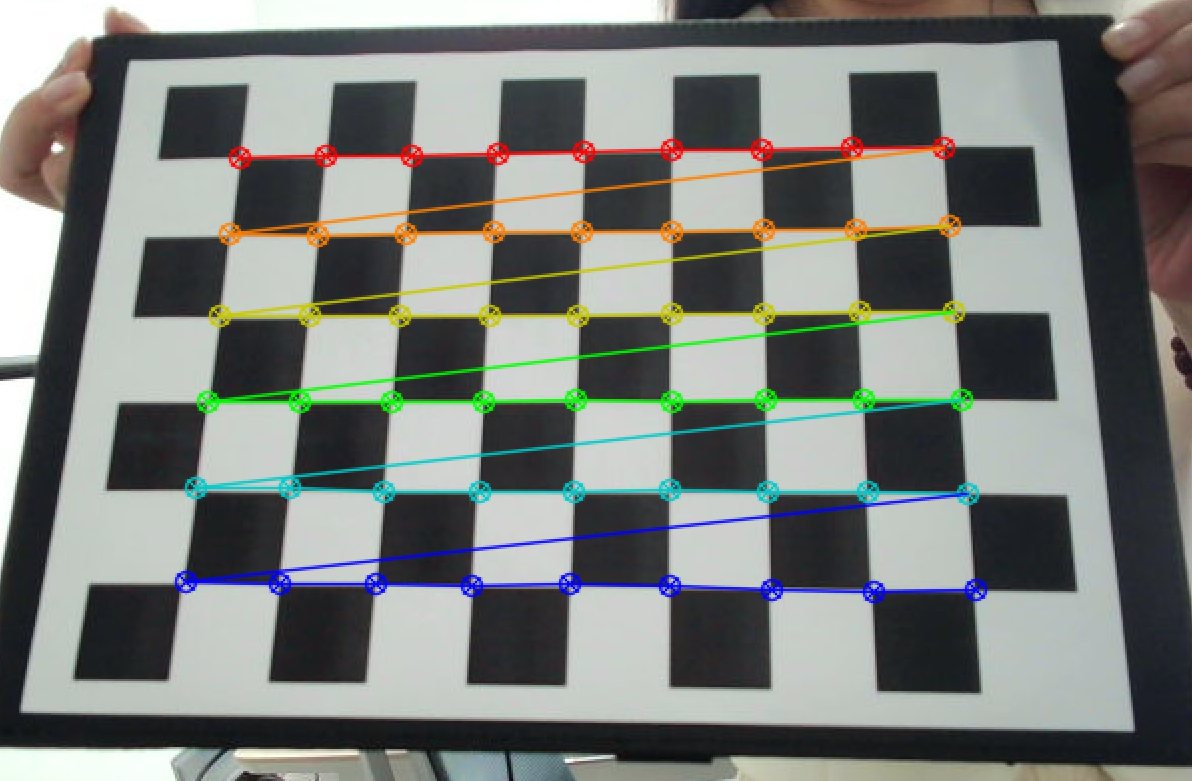}
\caption{\label{fig:chessboard} A planar checkerboard pattern for intrinsic and extrinsic camera calibration. Each camera was individually calibrated using 300 images of the calibration board captured from various angles and distances to ensure robust parameter estimation.}
\end{figure}



\subsection{Camera Calibration}
Figure \ref{fig:chessboard} shows a standard calibration procedure for each camera to estimate its intrinsic parameters (focal lengths, principal point) and lens distortion coefficients, aligning image coordinates with the true optical geometry. Our setup captures 300 images of a known calibration target checkerboard from various angles and distances, and the entire calibration process takes approximately 30 minutes. By processing these images, we solve for:
\begin{itemize}
    \item \textbf{Intrinsics:} 
          The camera matrix 
          $\mathbf{K}$, which contains focal lengths $(f_x, f_y)$ 
          and principal point $(c_x, c_y)$, 
          plus distortion coefficients $\boldsymbol{\kappa}$ 
          (covering both radial and tangential effects).
    \item \textbf{Reprojection Error:} 
          An average 2D pixel error, which indicates how well the calibrated 
          parameters explain the observed corners 
          of the known 3D calibration pattern.
\end{itemize}





\subsection{Augmented Reality Neuron Navigation System}
\begin{figure}[t]
    \centering
    \subfloat[3D Unity brain model.]{
    \includegraphics[width=0.95\linewidth]{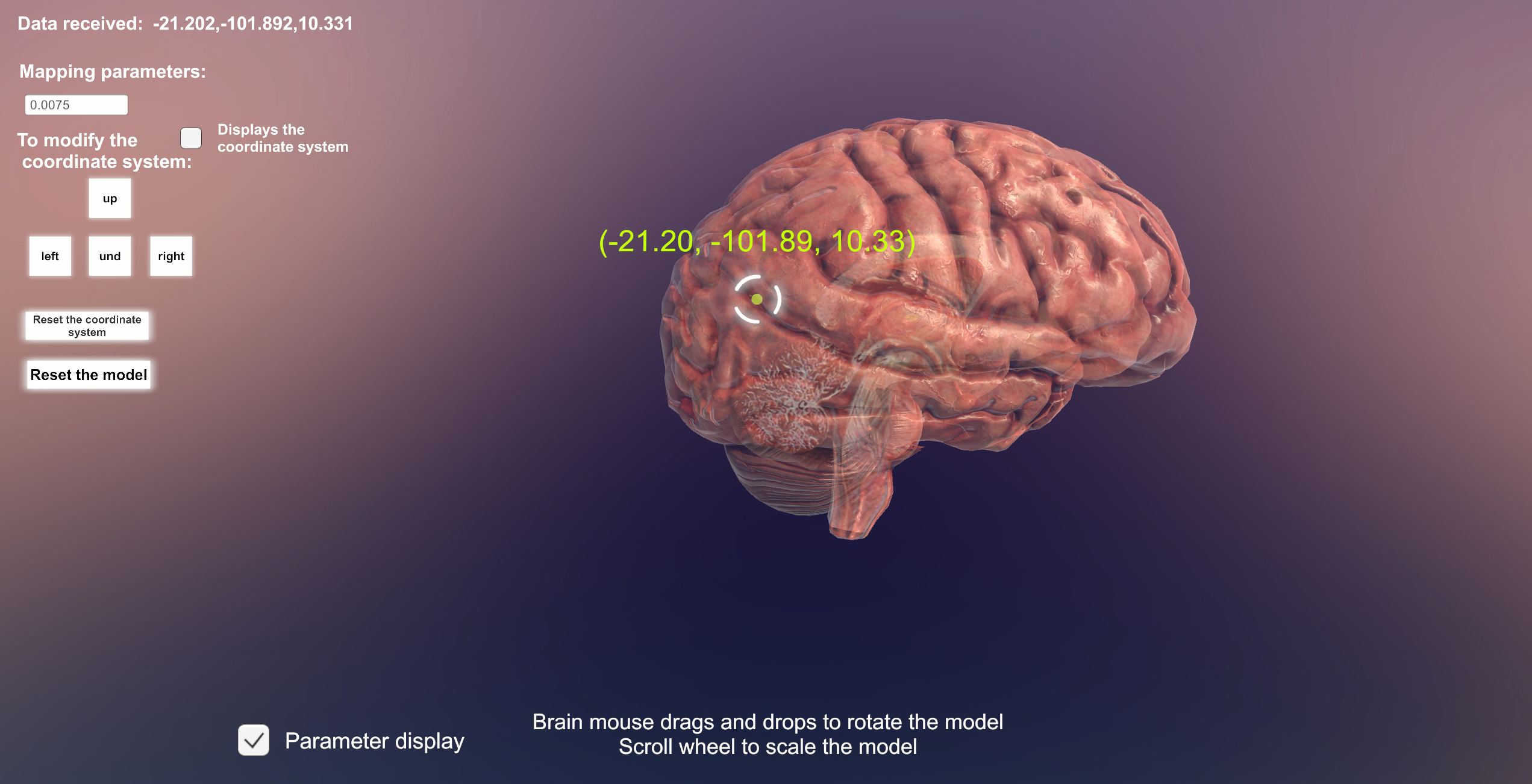}
    \label{fig:3D_brain_model}
    } 
    
    \subfloat[Augmented-reality neuronavigation.]{
    \includegraphics[width=0.95\linewidth]{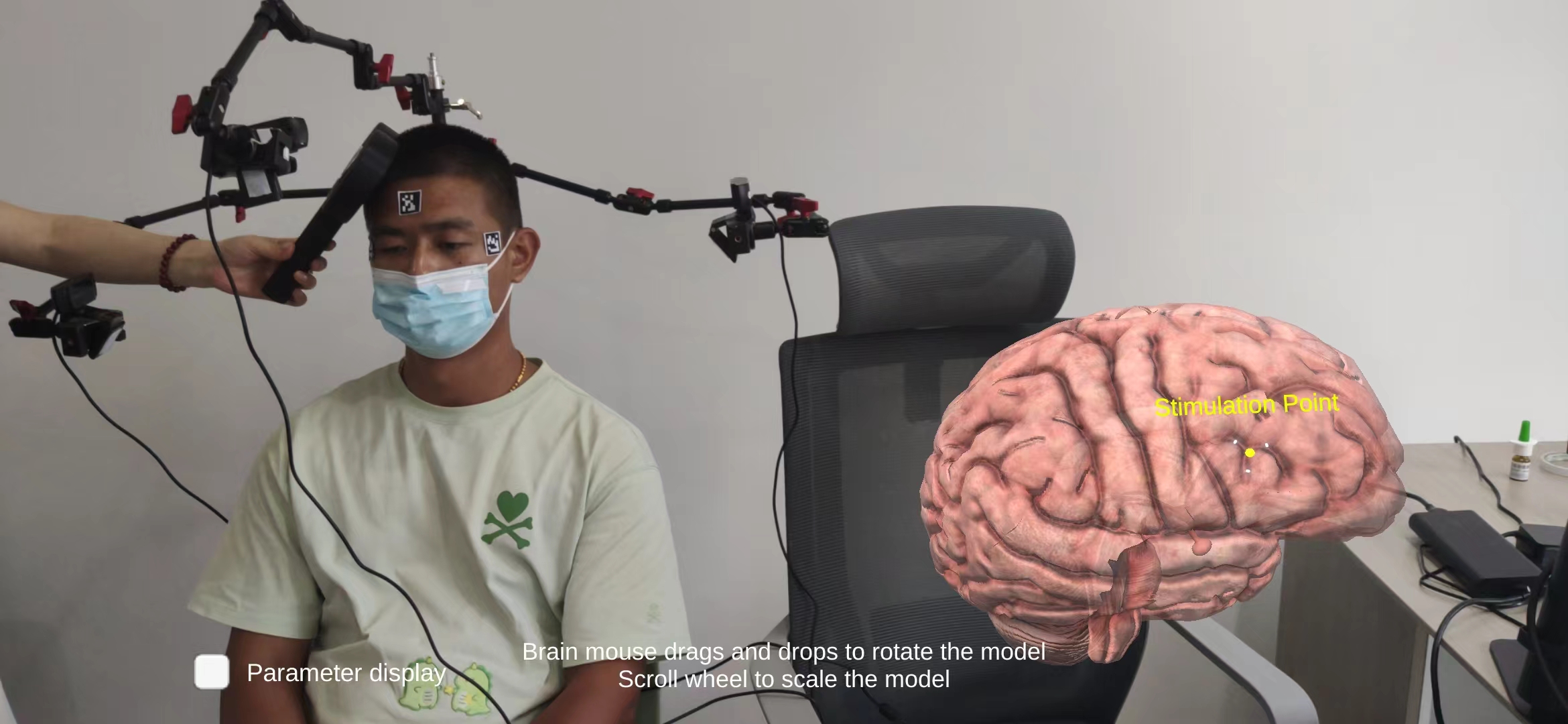}
    \label{fig:AR display}
    } 

    \subfloat[Augmented-reality neuronavigation.]{
    \includegraphics[width=0.95\linewidth]{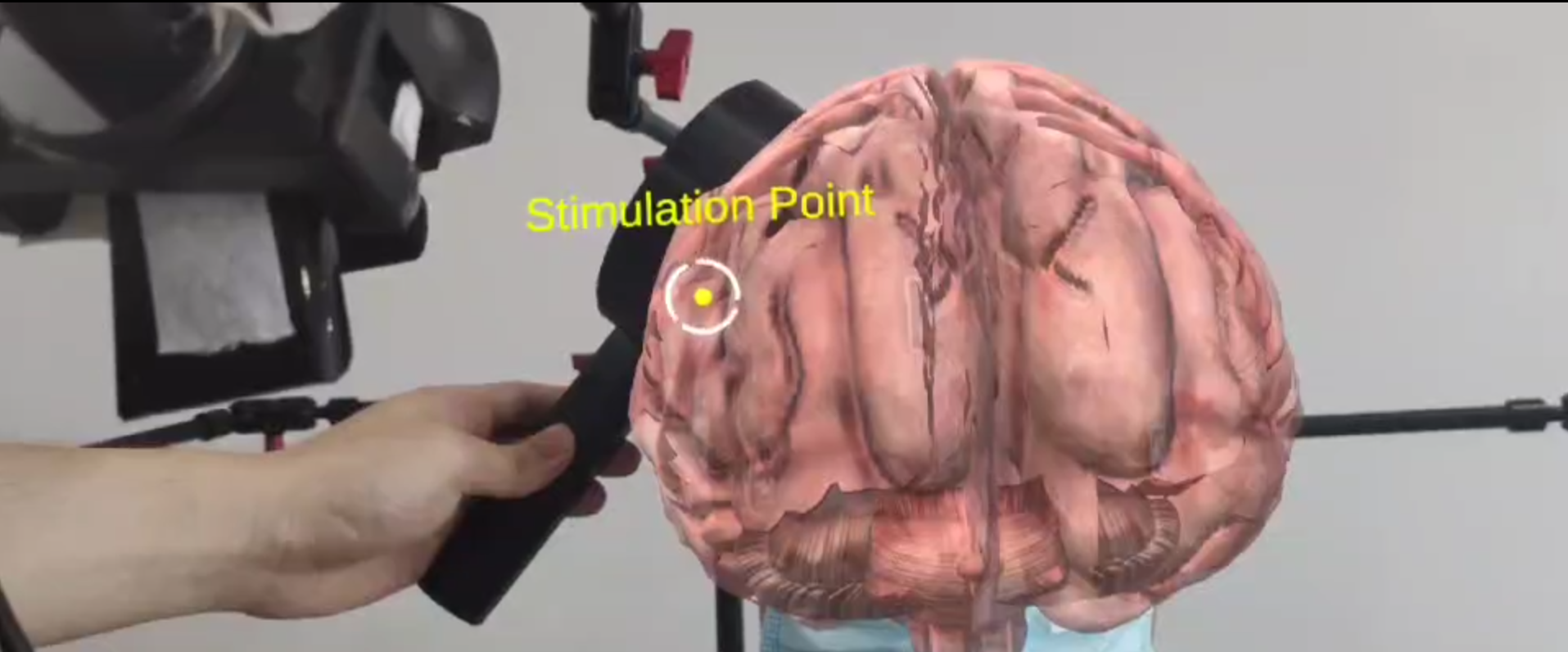}
    \label{fig:AR_display_overhead}
    } 
    
    \caption{Visualization of the neuronavigation system. (a) Real-time 3D visualization of the TMS stimulation point integrated with a digital brain model in Unity. (b) Augmented reality interface displaying the stimulation point. (c) Augmented reality visualization spatially registered and overlaid directly onto the patient’s head.}
    \label{fig:neuronavigation system display}
    \vspace{-0.3cm}
\end{figure}

To enhance the visualization of the targeted brain region during TMS-guided neuronavigation, we first develop a 3D brain model in Unity that dynamically updates to indicate the current stimulation target based on the real-time optical based multi-camera tracking system using the transmission control protocol‌ (TCP). The estimated simulation point is mapped onto the brain model, allowing clinicians to easily visualize the precise location of stimulation in a virtual environment, as shown in Figure~\ref{fig:3D_brain_model}.


We integrate AR technology to project the 3D brain model into the real-world environment. The augmentation improves usability in clinical settings and ensures a seamless transition between digital visualization and practical application. We use AR Foundation for real-time AR visualization. The system transmits tracking data from a PC local network to the Android system. This method ensures efficient updates and maintains responsiveness. AR headsets or mobile AR devices allow clinicians to view the brain model either shown alongside the patient or spatially registered and overlaid directly onto the patient’s head, and is scaled to match individual head dimensions. This feature improves spatial precision and simplifies operation. Figure \ref{fig:AR display} illustrates the AR neuronavigation. An Android mobile phone (Xiaomi Mi 9) serves as the visualization device. The phone has a 6.39-inch screen with a resolution of $2340 \times 1080$.

The AR neuronavigation system provides a direct and immersive visualization that supports intuitive TMS coil adjustments by allowing users to look straight onto the stimulation site, reducing reliance on abstract viewpoints and minimizing training demands associated with unnatural viewing angles. The integration of Unity-driven 3D modeling with real-time AR visualization should significantly enhance both accuracy and accessibility, making TMS neuronavigation more efficient and easy to use for medical practitioners.

\section{Results} 
\subsection{2D Reprojection Error}
\begin{figure}[t]
\centering
\includegraphics[width=1\linewidth]{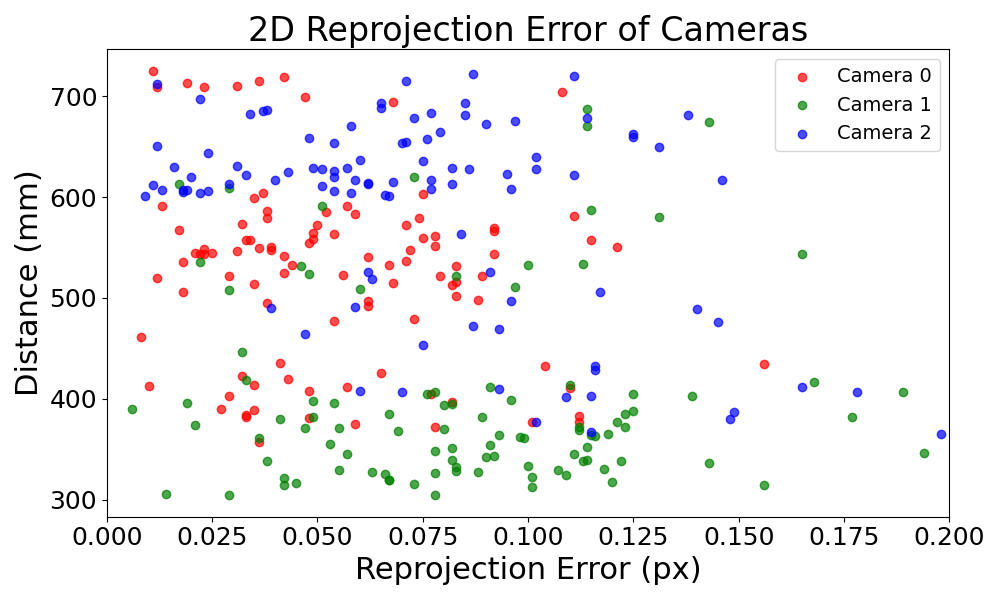}
\caption{\label{fig:2D projection error coordinate system} Relationship of re-projection error and distance across all three cameras. Each coloured cluster represents data captured from a different camera. The distribution of points suggests variations in accuracy and distance measurement across different camera perspectives.}
\end{figure}

The performance of the multi-camera tracking system is further evaluated using the 2D re-projection error, computed from 100 randomly sampled data points per camera (see Equation~\ref{eq:2D reprojection error}). This metric reflects the system’s accuracy in estimating tag positions in the image plane, serving as an indicator of calibration quality and overall tracking robustness. Figure~\ref{fig:2D projection error coordinate system} presents the relationship between re-projection error and tag-to-camera distance for all three cameras in the setup. Across distances ranging from 300~mm to 750~mm, the majority of data points remain well below 0.15~px. Camera 0 reports a mean error of approximately 0.063~px, Camera 1 about 0.064~px, and Camera 2 around 0.069~px, with variation between samples but no systematic increase in error as distance grows.

These results indicate that the intrinsic calibration and lens distortion compensation are effective in maintaining accuracy over a broad range of working distances. Importantly, the observed sub-pixel error values confirm that the corner detection process is highly precise and that the pose estimation pipeline provides a stable and reliable solution across all three cameras.

\begin{figure*}[tp]
\centering
\includegraphics[width=1\linewidth]{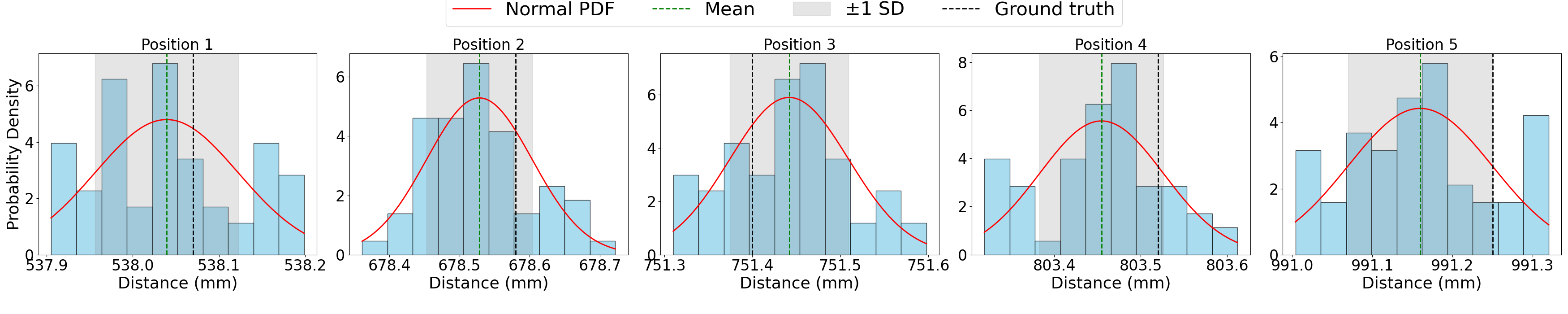}
\caption{\label{fig:distance_R&P}Probability distribution plots of the detected distance in the evaluation of the multi-camera neuronavigation system across five positions.}
\vspace{-0.3cm}
\end{figure*}

\begin{figure*}[tp]
\centering
\includegraphics[width=1\linewidth]{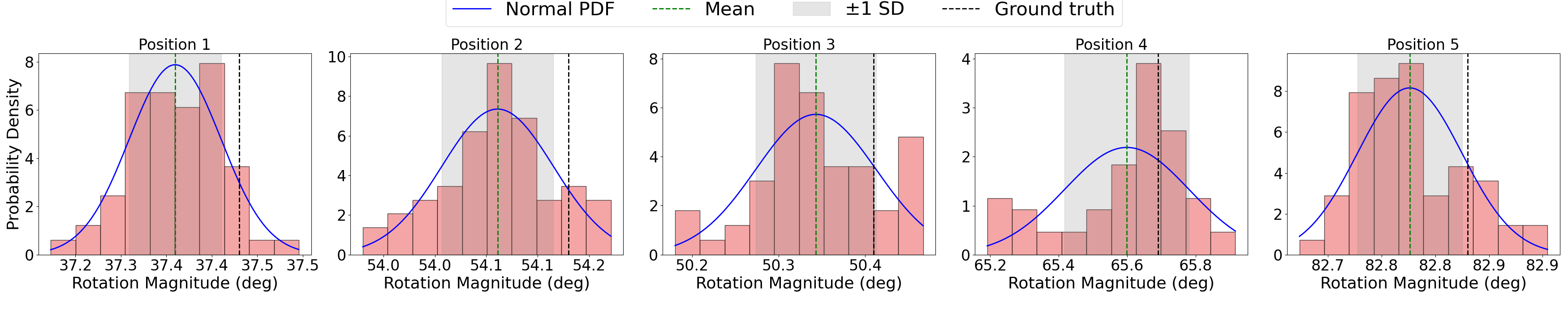}
\caption{\label{fig:angle_R&P}Probability distribution plots of the detected orientation in the evaluation of the multi-camera neuronavigation system across five positions.}
\vspace{-0.3cm}
\end{figure*}

\subsection{Precision and Accuracy}
The precision and accuracy experiment evaluates the system’s ability to provide stable measurements of the same tag target under identical conditions, while also quantifying the absolute error relative to ground truth. Precision reflects how small the variations are in repeated measurements, whereas accuracy indicates how close the measured values are to their true positions. The experiment was conducted by keeping the TMS coil stable and stationary while collecting 100 measurements at five different positions. This setup allows for a joint analysis of depth and angular performance.

Based on Figures~\ref{fig:distance_R&P} and \ref{fig:angle_R&P}, the analysis of precision and accuracy for the proposed multi-camera neuronavigation system indicates consistently narrow near-Gaussian distributions for good precision and accuracy. The histograms confirm that measurement variability remains very low. Comparisons with reference values show that absolute errors are confined to the millimeter range which support reliable accuracy over multiple trials.

For distance, the mean values remain stable across all five positions, ranging between 530~mm and 990~mm. Standard deviations are small, generally between 70~µm and 90~µm. For example, Position 1 shows a mean of 538.05~mm with a standard deviation of 0.08~mm, Position 3 records 751.5~mm ± 0.08~mm, and Position 4 centres at 803.51~mm ± 0.07~mm. These results confirm that depth estimation is both precise and robust, with only minor variations likely due to tag detection variability or subtle calibration residuals.The absolute errors compared to ground truth remained within one millimeter, indicating that depth estimation is both precise and accurate, with only minor deviations due to tag detection variability or residual calibration effects.

The rotation magnitude measurements likewise demonstrate excellent consistency, with standard deviations typically around 0.04–0.06°. Position 1 records a mean of 37.39° ± 0.05°, Position 2 shows 54.09° ± 0.05°, and Position~5 yields 82.84° ± 0.05°. The close Gaussian fits indicate angular noise is well-controlled, and deviations across positions are minimal, likely arising from minor calibration imperfections or marker placement differences. Absolute angular errors relative to the reference orientation remained well within 1°, demonstrating that the system not only minimises angular noise but also maintains strong alignment accuracy across different target positions.

Overall, the system achieves distance precision within 0.07–0.09~mm, angular precision within 0.04–0.06°, and absolute errors below 0.5~mm and 0.3° for distance and rotation respectively. These results highlight the robustness of the multi-camera setup, confirming that it delivers both high precision and accuracy across diverse experimental conditions.

\subsection{Position Acquisition Rate}
\begin{figure}[t]
\centering
\includegraphics[width=1\linewidth]{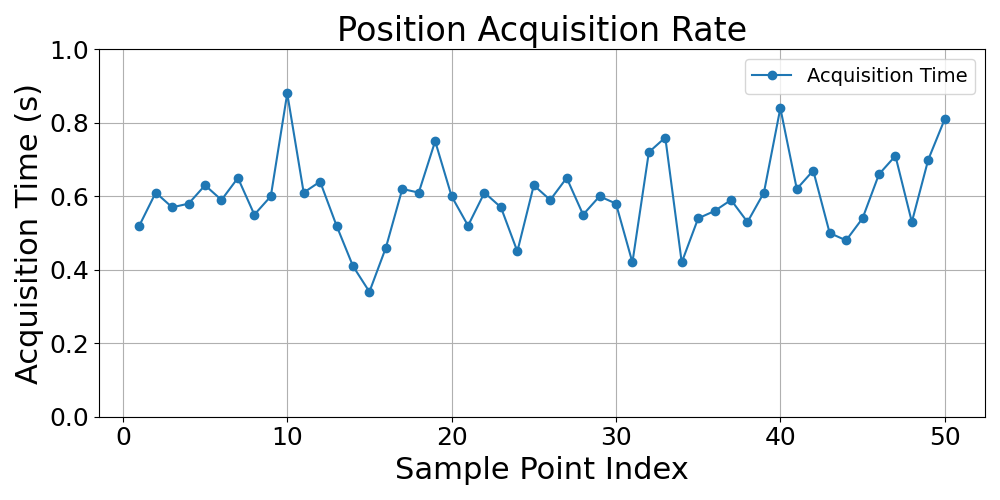}
\caption{\label{fig:position acquisition rate} Position acquisition times over 50 trials. The system consistently acquires positions in under one second, with most measurements below 0.8~seconds for interactive applications.}
\end{figure}

In positioning systems for neuronavigation, the position acquisition rate is a critical factor affecting real-time usability and user experience. Given the multi-camera setup used in our system, the computational demand can scale with the number of views, potentially introducing latency. To evaluate latency, we conduct empirical measurements on 50 consecutive frames, recording the system's end-to-end acquisition time for each localization cycle. Figure \ref{fig:position acquisition rate} illustrates real-time position acquisition analysis. The average acquisition time is 0.59~s with a standard deviation of 0.09~s, and 94\% of the trials completed under 0.8~s. The fastest response is 0.34~s, and the slowest is 0.88~s. These results show that the system is generally fast enough for interactive use.

\subsection{Stimulation Point Localization Accuracy}
\begin{figure}[t]
    \centering
    \subfloat[The positioning of 15 designated TMS stimulation targets for accuracy testing.]{
    \includegraphics[width=0.95\linewidth]{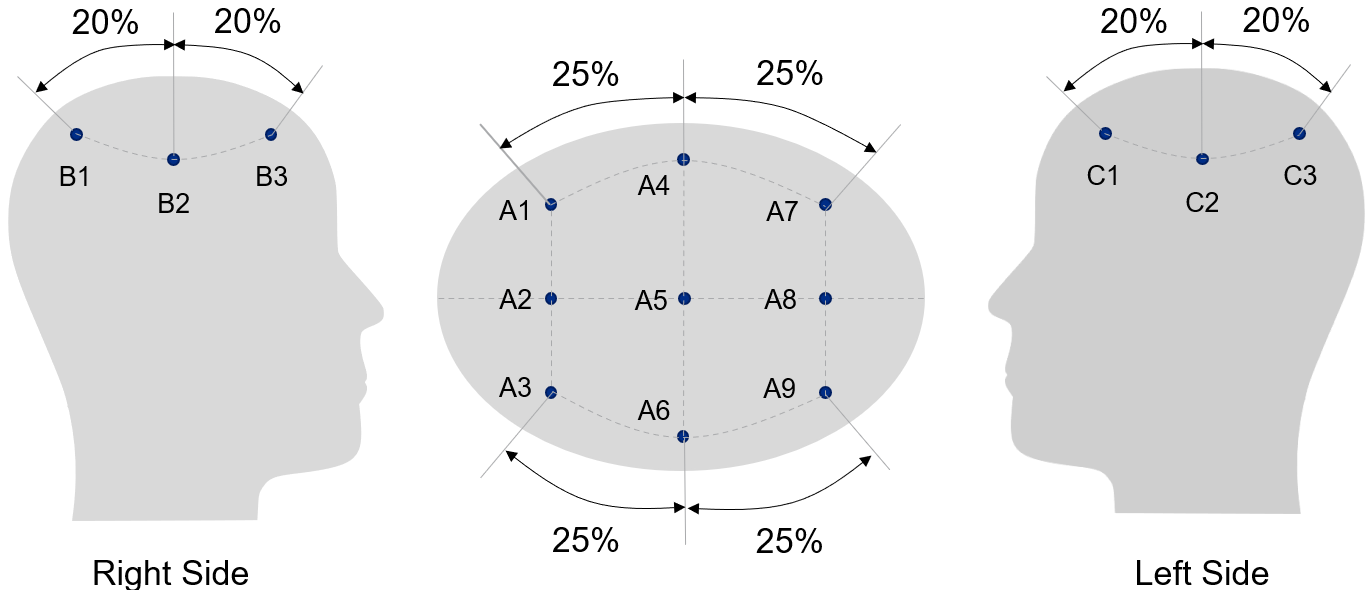}
    \label{fig:accuracy point location}
    } 
    
    \subfloat[Bar chart illustrating the stimulation target localization accuracy across 15 different test points (A1--A9, B1--B3, C1--C3). Each bar’s height represents the measured 3D re-projection error in millimeters.]{
    \includegraphics[width=0.95\linewidth]{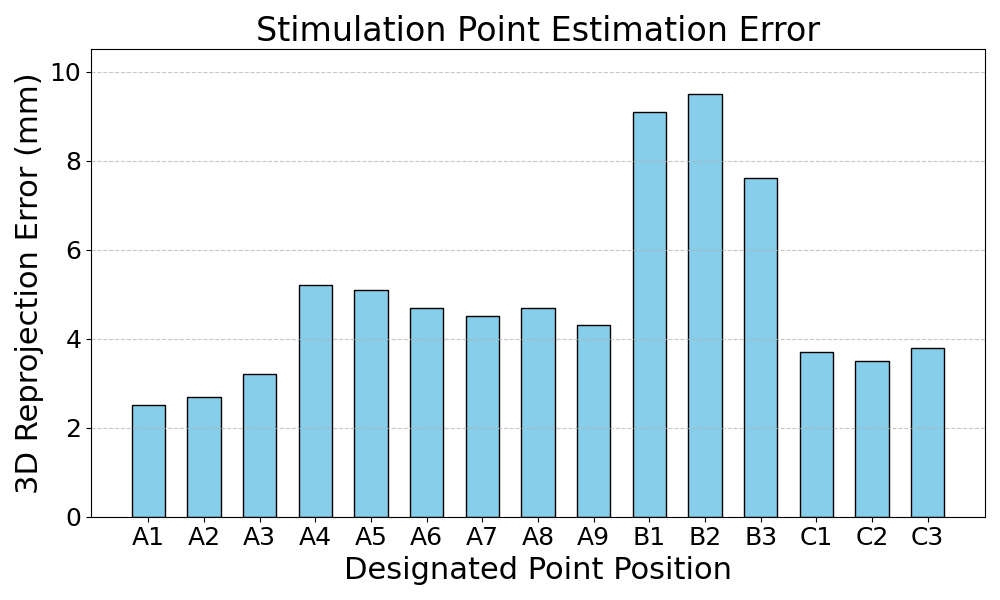}
    \label{fig:accuracy box plot}
    } 
    
    \caption{Selected stimulation target localization accuracy.}
    \label{fig:stimulation point localization accuracy}
    \vspace{-0.3cm}
\end{figure}

We define our system's accuracy through the Euclidean distance between the estimated 3D position, \(\hat{\mathbf{X}}\), and the known ground‐truth position, \(\mathbf{X}_{\text{true}}\). The ground‐truth positions are known locations measured by optical trackers. This metric effectively captures both the systematic bias (overall accuracy) and the standard deviation. Figure~\ref{fig:accuracy point location} shows the location of 15 selected points for accuracy testing and Figure~\ref{fig:accuracy box plot} illustrates localization accuracy across 15 different test points.

Figure~\ref{fig:accuracy box plot} reveals that the stimulation point estimation error varies across designated point positions. This variation in error indicates that there is both systematic bias and random variability in our system. 33.3\% of the measurements have errors below 4\,mm, 33.3\% of the measurements are between 4 and 6\,mm, and 16.7\% exceed 6\,mm. 

Figure~\ref{fig:accuracy SOTA box plot}  evaluates the proposed approach in comparison with three state-of-the-art neuronavigation systems. (1) Vuforia with HoloLens 1 (Vuforia HL1) uses marker-based tracking to overlay holographic guidance onto the surgical field. This system improves spatial awareness and assists with precise target localization~\cite{Frantz2018AugmentingMH}. (2) HoloLens-assisted ventriculostomy (HL ventriculostomy) integrates HoloLens and Vuforia to provide real-time visualization and interactive guidance. This system allows clinicians to align surgical tools with anatomical targets through a head-mounted augmented reality interface. The design enhances procedural accuracy and improves usability~\cite{Schneider2021AugmentedRV}. (3) Intel RealSense SR300 (Intel SR300) uses a hybrid tracking method that combines optical and depth sensing to estimate instrument position and orientation. The system fuses visual and spatial data to improve localization accuracy, particularly in real-time interventional procedures~\cite{asselin2018towards}.

Our method and Vuforia HL1 share lowest localization errors, with respetive average errors of 4.94~mm and 3.1~mm. Both systems use marker-based tracking, which tends to deliver stable and reliable performance. AprilTag MC benefits from the use of high-contrast fiducial markers that maintain accuracy across varied conditions, while Vuforia HL1 overlays AR with vision to assist with alignment. Despite their accuracy, both methods require consistent lighting and unobstructed marker visibility, which can pose limitations in real surgical environments. HL ventriculostomy causes a slightly higher average error of 5.2~mm. Holographic overlays, while useful for guidance, may introduce minor spatial distortions which degrade localization precision. Intel SR300 exhibits the highest error with an average of 20 mm. This method relies on depth sensing, which is inherently more susceptible to errors caused by lighting variation, surface reflectivity, and angular misalignment. These factors, especially when compounded near the periphery of the depth camera’s field of view, contribute to greater variability in measurements and lower overall accuracy.

\section{Case Study}
To assess the usability and accuracy of the proposed neuronavigation system in a realistic setup, we conducted a case study simulating TMS coil positioning over a designated cortical target.

\subsection{Participants}
We recruited ten participants through social media platforms (6 female, 4 male; mean age: 28.4 years, standard deviation: 3.9; min: 23; max: 35). All participants had no prior experience with TMS or the use of neuronavigation systems. Three participants reported prior exposure to AR immersive headsets and experience with Unity development. To ensure ecological validity, we deliberately recruited individuals with a medical background and basic knowledge of brain anatomy, reflecting the profile of the intended end-users of our proposed neuronavigation system, namely clinicians and researchers in neuroscience and neuromodulation.
\subsection{Procedure}
Before commencing the case study, we first provided participants with a brief introduction to TMS and the relevant aspects of basic brain anatomy. The overview included how TMS is used clinically and experimentally to non-invasively stimulate targeted cortical regions, as well as the importance of precise coil positioning for effective and reproducible stimulation outcomes.

For the case study, three to five stimulation targets were identified on the head, corresponding to these cortical regions. The participant was then instructed to align the TMS coil with each target using only the real-time AR visual feedback provided by our neuronavigation system. Once aligned, the system recorded the estimated coil position and orientation, which were compared against ground-truth values obtained through optical trackers. This allowed us to compute the 3D reprojection error for each trial to quantify positional accuracy. To assess repeatability and precision, each alignment was repeated five times at every target location, ensuring robust statistical evaluation of both depth and angular consistency.

\begin{figure}[t]
\centering
\includegraphics[width=1\linewidth]{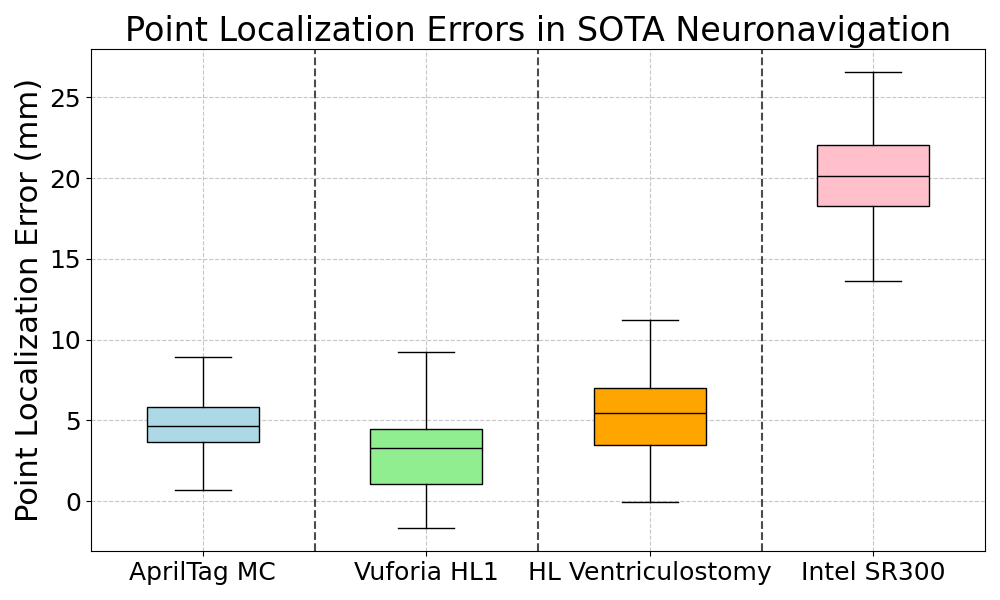}
\caption{\label{fig:accuracy SOTA box plot}Comparison of point localization errors between state-of-the-art and the presented neuronavigation systems. The boxplots represent the error distributions for each method with median, interquartile range, and overall spread of errors.}
\end{figure}

\subsection{Results}
\begin{figure}[t]
\centering
\includegraphics[width=1\linewidth]{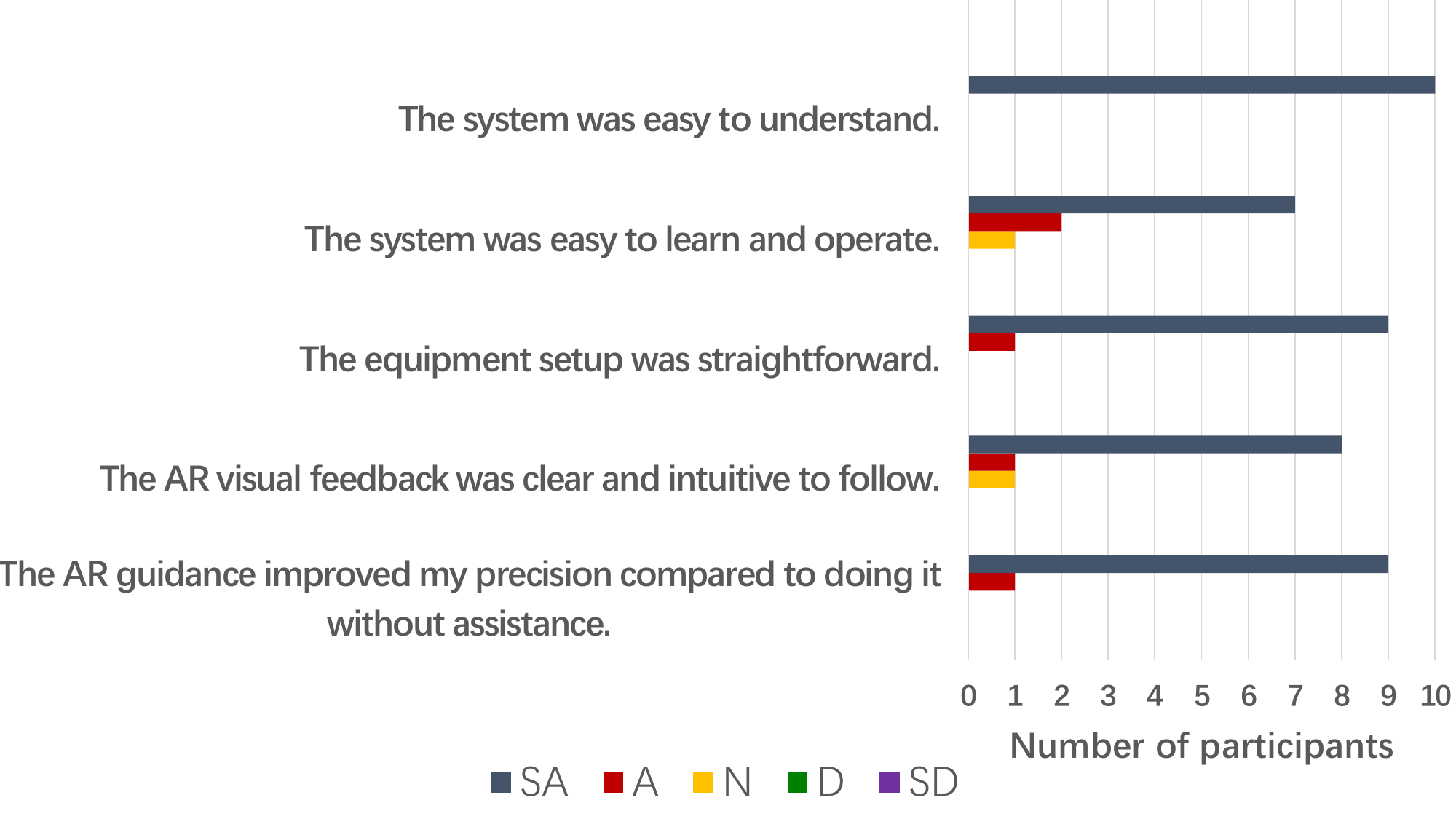}
\caption{\label{fig:questionnaire} Five-point Likert responses to the post-study questionnaire. Likert responses are colour-coded, and the scale shows the number of participants with the same rating. SA is strongly agree, A is agree, N is neutral, D is disagree, and SD is strongly disagree.}
\end{figure}

All participants successfully completed the case study by aligning the TMS coil with each of the designated target points. 
Their responses to the post-study questionnaire are summarized in Figure~\ref{fig:questionnaire}. In their questionnaire feedback, participants responded  positively about the system overall. All participants strongly agreed that the system was easy to understand, and 90\% agreed that the equipment setup straightforward. Similarly, the AR visual feedback was rated clear and easy to follow by all participants, with 80\% strongly agreeing and 20\% agreeing.

When asked about learning and operation, 90\% participants agreed that the system was easy to use, though 10\% reported minor initial difficulties. Thus, short training sessions or additional onboarding guidance could further reduce the learning curve. Importantly, 90\% of participants strongly agreed that AR guidance improved their precision compared to performing the task without assistance, confirming the system’s practical benefit in supporting coil alignment.

In addition to subjective usability ratings, objective accuracy measurements collected during the case study demonstrated that the system consistently achieved 3D reprojection errors within the millimeter range, with an average error below 6 mm across test positions. During the study, partial occlusion of some markers occurred as participants aimed to align the TMS coil with designated targets. Despite these occlusions, alignment results remained stable and within acceptable accuracy, underscoring the robustness of the system in realistic usage conditions.

Overall, the results highlight both subjective and objective strengths of the AR-based neuronavigation system. Participants found it easy to use and helpful for improving precision, while accuracy tests confirmed its reliability and robustness, even under partial marker occlusion. These outcomes demonstrate the system’s suitability for real-world neuronavigation tasks requiring consistent precision and user confidence.

\section{Discussion}

The results indicate that the concept can achieve high accuracy in projecting tag corners onto the image plane. As illustrated in Figure~\ref{fig:2D projection error coordinate system}, the narrow range of re-projection errors across varying distances and camera perspectives implies that the system remains largely unaffected by lens distortion or perspective-related issues. Minor variations in accuracy arise from changes in lighting, camera angles, or partial occlusion. Despite these factors, the consistent sub-pixel error levels highlight the precision of the optical tag detection process and validate the system’s effectiveness for real-time neuronavigation applications that require accurate spatial localization.

Our findings reveal that the high precision and accuracy observed in both distance and rotation angle measurements can be attributed to well-calibrated cameras with minimal lens distortion that ensure reliable depth estimation. The system benefits from a precise tracking algorithm and synchronized multi-camera setup that contributes to stable and consistent measurements (Figs.\! \ref{fig:distance_R&P} and \ref{fig:angle_R&P}). The data does not reveal a strong correlation between distance and error. Thus, the system compensates well for varying depths through accurate intrinsic calibration and lens distortion correction within the tested range. The small spread in errors across different distances and cameras suggests that the system is not significantly affected by perspective distortion or lens artifacts. Misalignments in optical tag detection and environmental factors can cause minor fluctuations in distance, such as the 0.1~mm variation in Position 1. The small variations in rotation angles may result from differences in tag orientation, or slight camera pose adjustments.

Figure~\ref{fig:accuracy box plot} highlights the high accuracy achieved by the proposed system. It reflects the strength of the calibration pipeline, where precise estimation of both intrinsic and extrinsic parameters ensures reliable reconstruction of 3D positions from 2D image observations. The rectangular markers from the April family contributes to robust and repeatable corner detection due to their high-contrast design and well-defined geometric structure. While pose estimation inherently depends on factors such as image noise, lighting conditions, and occlusion, the system demonstrates strong resilience under typical usage scenarios. The low error distributions suggest that the combination of accurate camera calibration and optical tag-based tracking forms a stable foundation for precise neuronavigation, even in the presence of real-world visual variability.

The comparison of the proposed system with state-of-the-art methods further highlights its competitive performance in localization accuracy, while it offers advantages in cost-efficiency, ease of deployment, and real-time AR visualization for clinical use. Figure~\ref{fig:accuracy SOTA box plot} indicates that localization accuracy remains comparable to conventional methods. In contrast to Vuforia with HoloLens1 and HoloLens-assisted ventriculostomy, this system does not require expensive AR headsets or complex marker calibration. The use of printed optical tags ensures straightforward implementation without additional hardware beyond standard cameras. The optical tag-based approach produces lower error variability than Intel RealSense SR300, which experiences greater fluctuations due to depth sensing inaccuracies. Depth-based tracking introduces inconsistencies caused by lighting conditions and sensor range limitations, whereas this system maintains stability for reliable target localization in neuronavigation applications (Figures~\ref{fig:distance_R&P}, \ref{fig:angle_R&P}, and \ref{fig:2D projection error coordinate system}).

The proposed system improves the ease of use and accessibility by solving the limitations of traditional optical neuronavigation systems, which require complex setups, dedicated workstations, and precise marker placements. In contrast, the optical tag-based approach leverages printed markers and standard cameras within a multi-camera framework for fast low-cost deployment and ease of use with minimal technical expertise. Its low-cost design eliminates the need for expensive optical tracking equipment, which makes it a practical and scalable solution for a wider range of institutions. Additionally, the integration of AR visualization enhances spatial awareness and user interaction by providing an easy to use and immersive interface for real-time navigation tasks. 

\section{Limitations and Future Work}
One limitation is that the system relies on clear marker visibility for at least some of the cameras. Lighting variations, or rapid movements however could impact tracking performance. The accuracy of multi-camera fusion also depends on proper calibration, and any misalignment could introduce errors.

We suggest future work focusing on refining marker detection algorithms to better handle lighting variations and enhancing real-time calibration for multi-camera setups. Accuracy in challenging environments can be further improved by integrating hybrid tracking approaches, such as fusing optical tag-based tracking with depth sensing. Additionally, Bayesian filters could help smooth noisy measurements and improve the consistency of both localization and tracking. The system could also offer neuronavigation hints to guide users toward optimal stimulation points during procedures, which can improve procedural efficiency for a more streamlined clinical workflow. Finally, we see value in future work carrying out further studies with end-users to gain a more nuanced understanding of the system’s overall usability, effectiveness, and impact on clinical practice. The current results motivate such in-depth follow-up studies in clinical settings, which we leave as future work.

\section{Conclusion}
This paper has presented an optical tag-based neuronavigation system as a cost-effective and accessible alternative to existing solutions. The system balances accuracy, affordability, and ease of use of low-weight printed black-and-white fiducial markers and standard cameras, which eliminates the need for expensive optical tracking equipment and complex calibration. The proposed method achieves localization accuracy comparable to state-of-the-art approaches with a simpler setup. The integration of AR visualization simplifies handling for users and gives more immediate spatial  awareness for a neuronavigation experience that is easier to learn and more efficient. The findings suggest that  optical tags are a viable solution for neuronavigation, particularly in resource-limited settings, and serves as a valuable enabling tool for the AR-based medical and neuronavigation research community.


\section*{Acknowledgments}
{The authors would like to express their gratitude to Siqi Miao and Kaiyi Chen for their assistance with the drawings.}

\bibliographystyle{IEEEtran}

\bibliography{template}
\end{document}